\documentclass[aps,pra,twocolumn, superscriptaddress,showpacs]{revtex4-1}
\usepackage[T1]{fontenc}
\usepackage[latin1]{inputenc}
\usepackage[dvips]{graphicx,color}
\usepackage{rotating}
\usepackage{bm}        % \boldsymbol (bold math)
\usepackage{amssymb}   % \mathbb     (blackboard bold)
\usepackage{amsmath}
\usepackage{braket}
\def\jcp#1#2#3{J.~Chem.~Phys.~{\bf #1},\ #2\ (#3)}
\def\cpl#1#2#3{Chem.~Phys.~Lett.~{\bf #1},\ #2\ (#3)}

\def\pra#1#2#3{Phys.~Rev.~A~{\bf #1},\ #2\ (#3)}
\def\prl#1#2#3{Phys.~Rev.~Lett.~{\bf #1},\ #2\ (#3)}

\def\ii{\text{i}}

\def\k1{k_1}
\def\k2{k_2}
\def\q1{q_1}
\def\q2{q_2}

\def\({\left (}
\def\){\right )}
\def\[{\left [}
\def\]{\right ]}

\def \MS{\mathcal{S}}

\newcommand{\beq}{\begin{equation}}
\newcommand{\eeq}{\end{equation}}
\begin{document}
\date{\today}
%\title{Rydberg atom detection of the temporal coherence of cosmic microwave background radiation}

\title{Coherent dynamics of Rydberg atoms in cosmic microwave background radiation}

\author{Timur V. Tscherbul}
\affiliation{Chemical Physics Theory Group, Department of
Chemistry, and Center for Quantum Information and Quantum Control,
University of Toronto, Toronto, Ontario, M5S 3H6, Canada}
\author{Paul Brumer}
\affiliation{Chemical Physics Theory Group, Department of
Chemistry, and Center for Quantum Information and Quantum Control,
University of Toronto, Toronto, Ontario, M5S 3H6, Canada}

\begin{abstract}

 Rydberg atoms excited by cold blackbody radiation are shown to display long-lived
  quantum coherences on timescales of tens of picoseconds. By solving non-Markovian  equations
  of motion with no free parameters we obtain the time evolution of the density matrix, and demonstrate
that the blackbody-induced temporal coherences manifest  as slowly decaying (100 ps)
quantum beats
in time-resolved fluorescence.
An  analytic  model   shows the dependence of the coherent dynamics on
 the energy splitting
between atomic eigenstates, transition dipole moments,
and  coherence time of the radiation. Experimental detection of the fluorescence signal
from a trapped ensemble of $10^8$ Rydberg atom is discussed,
but shown to be technically challenging at present, requiring
CMB amplification somewhat beyond current practice.

% of the radiation  novel approach to measuring radiation of fundamental importance in astrophysics.
% up  a wealth of new astrophysical applications.
%of research  to novel d impact astrophysical observations.
%  In particular, we  non-linear  that blackbody radiation can induce measurable coherent dynamics in Rydbergquantum coherent effects play a significant role this picture is limited in esse rate induced coherent dynamics in atoms and molecules.

\end{abstract}

%R
\maketitle

%!!! Add to conclusions --->  We show that this picture breaks down in significant ways on time scales comparable to the coherence time of the blackbody radiation.

\section{introduction}

The  interactions  of atoms and molecules with incoherent light
(such  as  blackbody  radiation,  BBR)  play  a central role in
research     fields     as     diverse     as    photosynthesis
\cite{Fleming,Scholes,ScholesJPCL,PNAS},          photovoltaics
\cite{Scully},    precision    spectroscopy   and   measurement
\cite{BBRshifts}, and atomic and molecular cooling and trapping
\cite{RA,Bas}.  Thermal  BBR  is  a  ubiquitous  perturber that
shifts  atomic  energy  levels  \cite{FarleyWing}, limiting the
accuracy of modern atomic clocks \cite{BBRshifts,Gibble}, and reducing
the  lifetime of Rydberg atoms \cite{RA,Kleppner82,6K,Tada} and
trapped   polar   molecules   \cite{Bas}.   Recent  theoretical
developments   suggest,  however,  that  quantum  noise-induced
coherence  effects  induced  by BBR can be used to cool quantum
systems  \cite{Cooling}  and  enhance  the  efficiency of solar
cells \cite{Scully}.

%    From a more fundamental viewpoint, the interaction of BBR with matter may be used to probe the Fundamental understanding of the response of atomic and molecular systems to incoherent light is thus of significant applied interest,  The need for such understanding has recently become manifestly clear in the context of The most ubiquitous form of  incoherent light, a blackbody radiation, BBR,

The dynamical response of a material system to incoherent light is
determined, among other factors, by the coherence time, a timescale over which
the phase relationship between the different frequency components
of the light source is maintained \cite{Loudon}. A
natural light source such as the Sun is well characterized as a
black body radiation (BBR) emitter with temperature $T = 5.6\times
10^{3}$ K, and extremely short coherence time of $\tau_c =
\hbar/kT\sim 1.3$ fs \cite{MehtaWolf,KanoWolf,Zaheen,Hoki,Leonardo}, where $k$ is the Boltzmann
constant. As a consequence, incoherent excitation of atomic
systems on timescales relatively long compared to
$\tau_c$ produces stationary mixtures of atomic eigenstates that do not evolve in time
\cite{PNAS,JB,Hoki,Leonardo}.
However, the coherence time of BBR
 increases with decreasing temperature and can reach values in excess of 2 ps at 2.7 K, the temperature of the cosmic microwave background radiation (CMB) \cite{CMBbook}.
This motivates interest in examining 
the temporal dynamics of atomic systems interacting with the CMB.  However, since the CMB intensity is $(300/2.7)^4$ =  $1.5\times 10^{8}$
weaker than that of BBR at 300~K,  the absorption signal in most ground-state atoms and molecules
even with suitably amplified CMB radiation \cite{Amplifiers}, is very small.
As  an initial step toward resolution of this difficulty, we propose to use highly excited Rydberg atoms, whose large transition dipole moments make
them extremely sensitive to external field perturbations \cite{RA}.  Previous experimental work has explored the absorption of BBR by Rydberg atoms,
leading to population redistribution, photoionization, and lifetime shortening \cite{RA,Kleppner82,6K}. However,  these experiments
were focused on measuring population dynamics with no attention  to        coherence
       effects. Similarly, no attention has been paid to coherence
       properties of CMB and the role it might play in enhancing
       cosmological information (e.g. \cite{Weinberg,Hinshaw,Amanullah,Hakim}).

%BBR is known to be responsible for shortening the lifetime of Rydberg atoms, and  a source of  the related lifetime shortering
% over the timescales of interest.

In this Article we examine  long-lived quantum coherence effects that 
occur in one-photon absorption of cold black body radiation (CBBR -- a term that we henceforth use to denote BBR at 2.7~K) by highly excited Rydberg atoms \cite{RA}. 
 Using a non-Markovian approach \cite{JB,TBP} to explore the dynamics of one-photon CBBR
 absorption, we show that the time-dependent fluorescence intensities of Rydberg atoms
 exhibit the quantum beats due to the coherences induced by a suddenly turned-on interaction with
 CBBR.
This suggests an experiment to explore the coherence 
properties of a cold trapped ensemble of Rb atoms in the presence
of CBBR.
  Our results demonstrate that
  non-Markovian and quantum coherence effects play a major role in short-time
  population dynamics induced by CBBR. 
  %Although not yet addressed   by the astrophysics community,   
%a measurement of such coherence properties for the
 % CMB   could provide additional insight into cosmological models of the
  %early Universe \cite{Weinberg,Hinshaw,Amanullah}.

Furthermore,  we develop  an analytical model for the coherences
in the long-time limit that is valid for an arbitrary noise source, here applied to CBBR.
The model  reproduces the coherent oscillations observed in numerical simulations
of the density matrix, and provides insight into the role of the energy level
splittings, transition dipole moments, and the coherence time of the radiation
in determining the time evolution of the coherences.  Significantly, we show that the ratio of coherences to populations declines with time as 1/$|\omega_{ij}t|$, where $\omega_{ij}$ is the energy splitting between the eigenstates $i$ and $j$.  Thus, the physical origin of the long-lived coherences is due to the small energy splittings between the eigenstates populated by  one-photon absorption of CBBR.

The paper is organized as follows.  Section II discusses the theory and Section III provides results and a discussion of the nature of the development and depletion of the coherences.

%and energy transfer in physics, chemistry, and  technology. On a more fundamental leve  and it has been suggested that quantum coherent effects can be used to enhance the efficiency of solar cells. Recently, it has been suggested that incoherent radiation can be used to cool atoms
%  and affects the performance  It is also of fundamental interest  currently attracting a much interest in a number of research fields. context of  range of fields, including precision measurement \cite{clocks}, atomic and molecular cooling and trapping \cite{Bas},  Rydberg atoms, and biophysics. In particular, ....EXPAND ON THE TOPIC.

\section{Theory}

Theoretically, the interaction of blackbody radiation with atoms
is usually considered within the framework of Markovian quantum
optical master equations \cite{BPbook}, leading to
Pauli-type rate equations for state populations parametrized by the
Einstein coefficients.  These treatments generally assume that the
coherences induced by BBR are negligibly small.  The non-Markovian approach adopted here \cite{JB,TBP,Leonardo} allows us to examine these
noise-induced coherences and memory effects arising from a finite correlation time of BBR.

 The time evolution of atomic populations and coherences under the influence
of incoherent radiation (such as BBR), suddenly turned on at $t =
0$, is given by \cite{Zaheen,JB, TBP}
\begin{multline}\label{EOM}
\rho_{ij}(t)  = \frac{\langle \mu_{i0}\mu_{j0}^* \rangle_p}{\hbar^2}  e^{-\ii\omega_{ij}t}  \\ \times \int_0^t d\tau' \int_0^t d\tau'' \mathcal{C}(\tau',\tau'') e^{\ii\omega_{i0}\tau'} e^{-\ii\omega_{j0}\tau''}
%\rho_{ij}(t)  = \frac{\langle \mu_{i0}\mu_{j0}^* \rangle_p}{\hbar^2}  e^{-\ii\omega_{ij}t} \int_0^t d\tau' \int_0^t d\tau'' \mathcal{C}(\tau',\tau'') e^{\ii\omega_{i0}\tau'} e^{-\ii\omega_{j0}\tau''}
\end{multline}
Here $\rho_{ij}(t)$ are the elements of the atom density matrix in the
energy representation, $\mu_{i0} = \langle 0 | \hat{\mu}
|i\rangle$ are the transition dipole moment matrix elements
connecting the initial atomic eigenstate $|0\rangle = |n_0 l_0
m_0\rangle$ and the final states $|i\rangle =  |n l m\rangle$ with
energies $\epsilon_0$ and $\epsilon_i$, $\langle ...\rangle_p$
denotes polarization-propagation average \cite{Griffiths}, and
$\omega_{ij}=(\epsilon_i-\epsilon_j)/\hbar$.
For the sake of clarity, we further assume that the atom resides
in a single state $|0\rangle = |n_0 l_0 m_0\rangle$ before the BBR
is turned on at $t=0$. Since $\rho_{00} \simeq 1$ at all times,
the density matrix [Eq. (\ref{EOM})]
 describes the populations and coherences among the states populated by BBR excluding the initial state \cite{JB}.
% and  we can exclude the initial state from the eigenstate basis.
% and is not included in Eq. (\ref{EOM}) because its population remains close to unity at all times.

% Mention towards the end: (1)  (\ref{EOM}) can be easily generalized to the case of several initial states (a Rydberg wavepacket),

The dynamics of the atom's response to incoherent radiation is
determined by the two-time  electric field correlation function
$\mathcal{C}(\tau',\tau'') = \langle \mathcal{E}(\tau')
\mathcal{E}^*(\tau'')\rangle$    in Eq. (\ref{EOM}). For a stationary
BBR source, the correlation function depends only on
 $\tau = \tau'-\tau''$, and is given by
\cite{MehtaWolf,KanoWolf}
\begin{equation}\label{C}
\mathcal{C}(\tau) = \mathcal{E}_0^2 ({90}/{\pi^4}) \zeta(4, 1+ \ii \lambda \tau)
\end{equation}
where $\zeta(4,x)$ is the generalized Riemann zeta-function
\cite{MehtaWolf,KanoWolf},  $\lambda = kT/\hbar$, $T$ is the temperature of
the BBR, and $\mathcal{E}_0^2 = [2\pi^3/(45 \hbar^3 c^3)](kT)^4 $ is the
mean intensity of the BBR electric field \cite{Itano,MehtaWolf,KanoWolf}.
Note that Eq. (\ref{C}) applies when $\omega_{i0}>0$ (absorption);
 $\mathcal{C}^*(\tau)$ should be used for stimulated emission 
($\omega_{i0}<0$). Because $\langle \mathcal{E}(\tau')
\mathcal{E}(\tau'')\rangle=0$ for CBBR \cite{Loudon,BPbook}, there is no
coherence between those levels populated in absorption and those
levels populated in stimulated emission from a given initial state
\cite{TBP}. Combining Eq. (\ref{C}) with Eq. (\ref{EOM}), and
evaluating the time integrals, gives (see Appendix A for details)
\begin{align}\label{pop}\notag
\rho_{ii}(t) = \frac{\langle  |\mu_{i0}|^2 \rangle_p}{\hbar^2} \biggl{(} & t [\mathcal{K}_0^{(+)} (\omega_{i0}, t) + \mathcal{K}_0^{(-)}(\omega_{i0}, t)] \\ &\,\,\,\, - \mathcal{K}_1^{(+)} (\omega_{i0},t) - \mathcal{K}_1^{(-)} (\omega_{i0},t) \biggr{)}
\end{align}
where
\begin{equation}\label{K}
\mathcal{K}_n^{(\pm)} (\omega, t) = \int_0^t  \tau^n \mathcal{C}(\pm \tau) e^{\pm \ii\omega \tau}d\tau
\end{equation}
 are half-Fourier transforms of $\tau$-scaled time correlation functions.
 In the long-time limit ($t\to \infty$), the right-hand side of Eq.~(\ref{pop}) grows
 linearly with $t$. Note that since we neglect spontaneous emission, the long-time limit
 is restricted to timescales short compared to the (very long)
 radiative lifetime, 200 $\mu$s, of the $65s$ state \cite{BeterovPRA}.
%Specifically, dropping the last two terms (which are negligibly small at large $t$) and
 Using an integral representation for the generalized Riemann zeta function, we obtain the limit
 (See Appendix A for details)
\begin{equation}\label{pop_inf}
\rho_{ii}(t) =  \frac{2\pi}{\hbar^2} \langle |\mu_{i0}|^2\rangle_p I(\omega_{i0}) t \quad (t\to \infty),
\end{equation}
where
$I(\omega)=\frac{2\hbar^3}{\pi c^3} \frac{\omega^3}{e^{\hbar\omega/kT}-1}$ is proportional to Planck's spectral density of BBR \cite{Loudon}.  Hence,  in the long-time (Markovian) limit, this
approach reduces to Fermi's Golden Rule \cite{Griffiths} commonly
used to calculate the rates of BBR-induced population transfer
\cite{RA,BeterovPRA}.

The off-diagonal elements of the density matrix are obtained in Appendix A as
\begin{align}\label{coh}\notag
\rho_{ij}(t) &= \frac{\langle \mu_{i0}\mu_{j0}^* \rangle_p}{\hbar^2}  \frac{1}{\ii\omega_{ij}} \biggl{(}   [\mathcal{K}_0^{(+)} (\omega_{j0}, t) + \mathcal{K}_0^{(-)}(\omega_{i0}, t)] \\ &\,\,\,\,
 - e^{-\ii\omega_{ij}t} [\mathcal{K}_0^{(+)} (\omega_{i0},t) + \mathcal{K}_0^{(-)} (\omega_{j0},t)] \biggr{)}
\end{align}
Note that due to the double half-Fourier transforms in Eq. (\ref{EOM}),
Eq. (\ref{coh}) is sensitive to frequency cross correlations in the CBBR.

      \begin{figure}[t]
    \centering
    \includegraphics[width=0.48\textwidth, trim = 0 0 0 0]{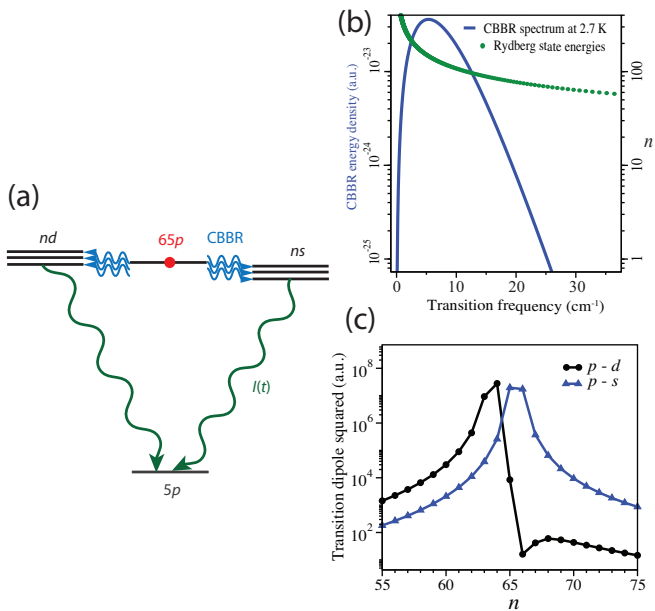}

    \renewcommand{\figurename}{Fig.}
    \caption{(a) Proposed experimental setup for observing long-lived quantum
     coherences with Rydberg atoms. At time $t=0$, the atom in the $65p$ Rydberg
     state (red circle) begins to interact with CBBR (wavy lines),
     leading to a decohering Rydberg wavepacket composed of the $ns$ and $nd$ states.
     The wavepacket evolves and decays to the $5p$ state, with the quantum coherences
     leaving their signatures in the fluorescence signal $I(t)$ (see text).
      (b) Binding energies of highly excited $ns$ Rydberg states of  $^{85}$Rb  together with the 2.7 K Planck
      spectrum of CBBR radiation. The zero of energy corresponds to the ionization threshold.
      (c) $n$ dependence of the calculated transition dipole moments squared from the initial $65p$ state to the  $ns$ states (triangles) and $nd$ states (circles) }\label{fig:picture}
\end{figure}

\section{Results and Discussion}

\subsection{Quantum dynamics of Rydberg atoms in CBBR: Populations and Coherences}

We now apply the approach developed in Sec. II to examine the effects of quantum
coherence in CBBR excitation of high$-n$ Rydberg atoms. In order to parametrize the equations of motion (\ref{EOM}),  the Rydberg energies and transition dipole moments for $^{85}$Rb  are calculated by solving the radial Schr{\"o}dinger equation for the Rydberg electron using the Numerov method \cite{RA,Zimmerman,TBP}. To verify the accuracy of our results, we calculated the spontaneous emission rates from the $30s$ Rydberg state to various final $np$ states.
These results agree with those reported in \cite{BeterovPRA} to within  $<5\%$.

Figure 1(a) shows the proposed setup for examining CBBR-induced
coherences. A highly excited Rydberg state of an alkali-metal atom
(here we focus on the $65p$ state of $^{85}$Rb)
is created at $t=0$ by e.g., excitation from the ground $5s$ state  \cite{RydbergWP}. The newly prepared
Rydberg state immediately starts to interact with the 2.7~K CBBR
background, establishing a coherent superposition of  the
neighboring $ns$ and $nd$ Rydberg states \cite{Kleppner82}.
In order to map out the time evolution of Rydberg populations and
coherences, Eq. (\ref{EOM}) is parametrized by the
accurate transition dipole moments of $^{85}$Rb and by the CBBR
correlation function given by Eq.~(\ref{C}).

 The rapid turn-on of CBBR acts as a coherent perturbation, creating a Rydberg
wavepacket that evolves with time, and then slowly decoheres.
Figure 1(b) shows the Rydberg energy levels of $^{85}$Rb superimposed on the CBBR spectrum at 2.7 K.  While the spectral width of
  the radiation is broad enough to excite the Rydberg levels with principal quantum numbers $n=35-115$,
  the transition dipole moments (Fig. 1c) decrease dramatically with increasing $\Delta n = n-n_0$, so most of the population
   transfer from the $65p$ state occurs to the neighboring Rydberg states with the largest transition dipole moments (see Fig. 1c) via one-photon absorption ($66s,\,64d$)
    and stimulated emission ($65s,\, 63d$). For this reason,   CBBR-induced photoionization
     occurs at a slow rate and  can be neglected  for $n_0=65$.  Spontaneous emission from the $65p$
     state is also neglected since it
      occurs on a much longer timescale (200 $\mu$s \cite{BeterovPRA}) than considered in this work.

\begin{figure}[t]
    \centering
    \includegraphics[width=0.4\textwidth, trim = 0 0 0 0]{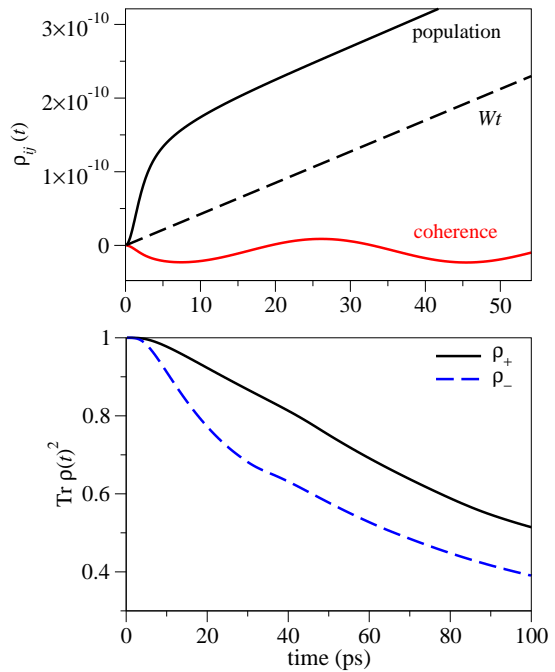}
    \renewcommand{\figurename}{Fig.}
    \caption{(a) The population of the representative $65s$ state of Rb as a function of time; (b) The time dependence of the purities for the absorption and stimulated emission blocks of the density matrix (\ref{purity}). The dashed line in the upper panel shows the expected Markovian behavior of the populations.}\label{fig:rho}
\end{figure}

Figure 2(a) shows the time evolution of several representative
density matrix elements given by Eqs. (\ref{pop}) and (\ref{coh}).
At $t \le$ 50 ps, the off-diagonal elements of the density matrix
are of the same order of magnitude as the diagonal elements, suggesting
the presence of  coherences that  play a  role in the
dynamical  evolution of a Rydberg atom during the first 50 ps of
its exposure to CBBR.
At short times, state populations exhibit substantial
deviations from the linear behavior predicted based on the
standard Markovian quantum optical master equation \cite{BPbook}.
The latter is shown in Fig.~2(a) as the  linear solution
$\rho_{ii}(t) = W_{0\to i}t$, where  $W_{0\to i}$ is the standard BBR-induced
transition rate  related to the Einstein $B$-coefficient \cite{RA}. The
exact non-Markovian population dynamics  is different in character
and magnitude \cite{TBP} but becomes linear in the larger $t$ limit.
%  familiar Markovian rate equations (based on the  approximation). These populations do not grow linearly with time  a linear

As shown in Fig. 2(a), the diagonal elements of the density matrix
grow linearly with time while off-diagonal elements oscillate.  As a result,  the populations begin to
significantly dominate over the coherences. Thus, BBR excitation produces a
stationary mixture of atomic eigenstates, with coherences playing
a negligible role in the long-time limit (nanoseconds) \cite{Leonardo,Zaheen}.
This gradual reduction of the coherences to population ratio is the mechanism of 
BBR-induced decoherence for the particular initial state.  
It differs from other cases \cite{Elran1,Elran2} where the initial state is a coherent superposition of energy eigenstates.

To see the decoherence times more clearly,  Fig. 2(b) shows a useful measure of
 decoherence---the purity of the density matrix
\cite{Schlosshauer}
\begin{equation}\label{purity}
\varsigma = \text{Tr}\left(\rho_{\pm}^2\right) = [N_\pm(t)]^{-1} \sum_{i,j=1} | \langle i | \rho_{\pm}(t) |j\rangle |^2
\end{equation}
where $\rho_\pm$ are the subblocks of the full density matrix
composed of the states populated in absorption and stimulated
emission from the initial state and the normalization factors $N_\pm(t) = \sum_i \langle
i|\rho_{\pm}(t) |i\rangle^2$ ensure trace conservation \cite{TG}.
The purity decays over a time scale  $ > $100 ps, which signals
the formation of an incoherent statistical mixture of atomic
eigenstates in the process of CBBR excitation.

%($\rho$ factorizes into two subblocks for chaotic light excitation, see above). Note that the sum (\ref{purity}) does not include the initial state $|0\rangle$ and the matrix elements on the right-hand side are renormalized so as to ensure that $\sum_i = 1^N$ only the excited-states of the density matrix In order to \cite{TG} decay of coherence

%$\mathcal{K}_2^{(\pm)} (\omega,t) = \int_0^t \tau \mathcal{C}(\pm\tau) e^{\pm i\omega \tau}d\tau$.

% ($n$, $l$, and $m$ stand for the principal, orbital, and azimuthal quantum numbers).
%(for the sake of simplicity, we assume that the atom is initially in a single eigenstate $|0\rangle = |n_0 l_0 m_0\rangle$ and the BBR is turned on at $t=0$).

As is typical of direct CBBR measurements, the populations in Fig. 2(a)
are quite small.  As such, we note standard CMB  amplification
practices \cite{Amplifiers}, which at present can give power gains in excess of
65 dB. Below we report results for a gain of 90 dB, which is technically possible,
but experimentally challenging.

\subsection{Observables: Time-resolved Fluorescence}

While clearly suggesting the existence of long-lived coherences on timescales of up to $\sim$100 ps,
neither  the density matrix elements nor the purity plotted in Fig. 2 are
experimental observables. To explore the possibility
of experimentally measuring the long-lived Rydberg coherences, we
evaluate the time-resolved fluorescence signal from the $ns$ and
$nd$ states of $^{85}$Rb populated by the interaction with CBBR
(see Fig. 1). These states decay to the $5p$ state of Rb
($|i_f\rangle$) by emitting a photon at a transition frequency of 620 nm, which can be
detected with high quantum efficiency. The total power emitted on
these transitions by  $N_a$ atoms is given by \cite{Demtroder,JB}
\begin{equation}\label{Ifluo}
I(t) = I_0 \text{Tr} \left\{ |\hat{\mu} i_f \rangle \langle i_f \hat{\mu}| \rho(t) \right\}
=I_0\sum_{i,j=1} \mu_{i i_f} \mu_{i_fj} \rho_{ji}(t)
\end{equation}
where $I_0=N_a \frac{4}{3} \omega^4/(4\pi \epsilon_0 c^3)$, $\epsilon_0$ is the vacuum permittivity,
 $c$ is the speed of light, and $\omega$ is the transition frequency, assumed  the same for all $i,j$ states (since $|\omega_{ij}|\ll |\omega_{ii_f}|$).

%\textbf{We  find  that  the  emission  signal  for  $N_a=10^8$	
%Rydberg  atoms  is  on  the  order  of  10$^{-17}$ W, or about
%$2\times  10^{-8}$  photons  in  the  first  10  ps,  which is
%hardly  detectable.}

\begin{figure}[t]
    \centering
    \includegraphics[width=0.4\textwidth, trim = 0 0 0 0]{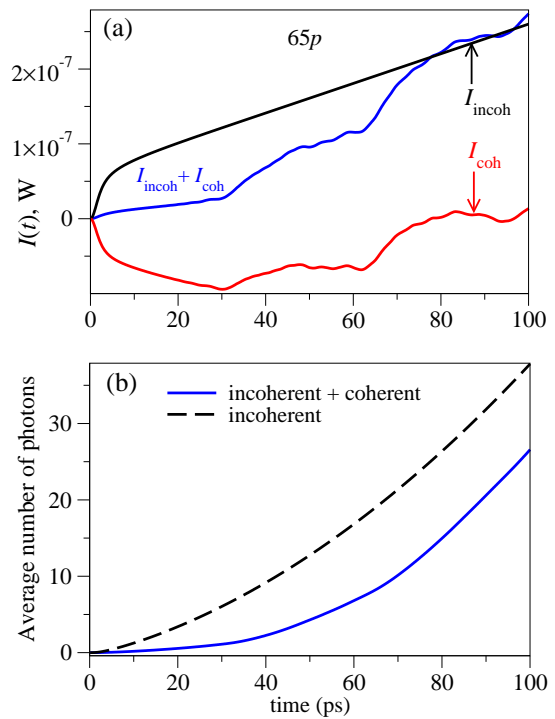}
    \renewcommand{\figurename}{Fig.}
    \caption{(a) Time-dependent fluorescence intensity for $N_a=10^8$ Rydberg atoms initially in the $n_0=65p$ state interacting with CBBR amplified by a
    factor of 90 dB. (b)  Average number of emitted photons $N_\text{ph}(t)$  (see text).
    The final state to which fluorescence occurs is $|i_f\rangle = |5p\rangle$. Also shown are the incoherent and coherent contributions to the total fluorescence intensity and to $N_\text{ph}(t)$. }\label{fig:rho}
\end{figure}

Figure  3(a) shows the calculated time dependence of the
fluorescence intensity for $N_a=10^8$ Rb atoms interacting with
amplified CBBR. The
time-resolved  emission signal displays pronounced oscillations
over  the timescales of 100 ps. The oscillations
can be  separated into  coherent
and    incoherent    parts,   $I(t)   =   I_\text{incoh}(t)   +
I_\text{coh}(t)$, with \cite{JB}
\begin{align}\label{separation}\notag
 I_\text{incoh}(t) &= I_0\sum_{i=1} \mu_{i i_f} \mu_{i_f i} \rho_{ii}(t)\\
 I_\text{coh}(t) &= I_0\sum_{i\ne j} \mu_{i i_f} \mu_{i_fj} \rho_{ji}(t);
\end{align}
The incoherent contribution $I_\text{incoh}(t)$ depends on 
the diagonal elements of the density matrix (populations) while
the coherent contribution $I_\text{coh}(t)$ specifically
highlights the role of quantum coherences. As shown in Fig.~3(a),
the coherent contribution to $I(t)$ remains significant up until
$t<100$ ps, suggesting the possibility of experimental observation
of CBBR-induced Rydberg coherences and their subsequent decoherence.

Figure  3(b) displays the time dependence of the
integrated fluorescence signal $F(t)=\int_0^t I(\tau)d\tau$
with $I(\tau)$ given by Eq. (\ref{Ifluo}), which represents the
experimentally measurable average number of photons emitted within the
time  window  $[0,t]$:  $N_\text{ph}(t)=F(t)/\hbar\omega$.
%{\bf We present an analytical model for the noise-induced coherences,
%which brings out the role of the energy splitting between the atomic eigenstates, the transition dipole moments, and the coherence time of the radiation as they determine the coherent response of an atomic system coupled to incoherent radiation. Experimental detection of the fluorescence signal is demonstrated to be technically challenging, requiring a cold trapped ensemble of  $10^{8}$ Rydberg atoms and low-noise, broadband amplification of the incident cold blackbody radiation by a factor of $10^9$.}
The calculated photon flux is $\sim$0.2 photons in the first 10 ps,
$\sim$2.3 photons in the first 40 ps, and $\sim$26.6 photons in
the first 100 ps of observation, assuming 100\% photodetection
quantum efficiency.
While not showing any coherent
oscillations, the integrated signal including the coherence
contributions [full line in Fig. 3(b)] is smaller than its incoherent
counterpart [dashed line in Fig. 3(b)] by a factor of 4 at $t=40$ ps and by 40\% at $t=
100$ ps. This difference represents a clear signature of
time evolution of the CBBR-induced coherences.

%\textbf{TIMUR: With respect to the next paragraph. It says
%that `` Coherent
%properties  of  the  CBBR  are  embedded  in  the  non-Markovian
%behavior of populations". I am not sure how these properties are
%related. Do you know?}

%and note that no for  detectable, number is about 5-10 times larger than typical experimental atom numbers this is only a factor of 10, which can be b is still vastly better than what we had with fluorescence detection.

%Figure 3(b) displays the results obtained for the initial state of $s$-symmetry (65$s$), for which the coherent contribution to the fluorescence signal is less pronounced than for the 65$p$ initial state. This difference highlights the role of the initial state in coherent dynamics, which is determined by the relative magnitude of the transition dipole matrix elements between the Rydberg levels populated by CBBR [see Fig.~1(c)] and the spontaneous emission matrix elements $\mu_{ii_f}$ in Eq.~(\ref{Ifluo}).

% spontaneous transitions to the 5$p$ state). whiThe time dependence of the fluorescence intensity is determined by a combination of the transition dipole moments, which exhibit oscillatory bahavior nature of the fluorescence

\begin{figure}[t]
    \centering
    \includegraphics[width=0.4\textwidth, trim = 0 0 0 0]{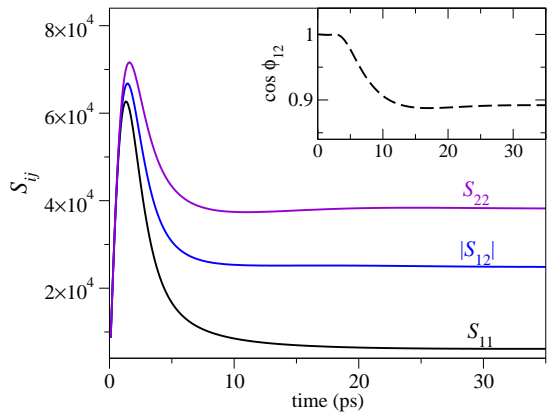}
    \renewcommand{\figurename}{Fig.}
    \caption{Time dependence of the coefficients $\MS_{ij}$. The inset shows the cosine of the phase angle $\phi_{12}$ as a function of time. Note that the $\MS_{ij}$ tend to constant values in the limit $t\gg \tau_c=2.8$ ps as appropriate for CMB.}\label{fig:Sij}
\end{figure}

\subsection{Analytics of noise-induced coherences and timescale for eigenstate formation}

As   shown   in  Figs.  2  and  3,  the  CBBR-induced  coherent
oscillations  survive  on  a  timescale  much longer ($\sim$100
picoseconds)   than   the  coherence  time  of  CBBR  at  2.7~K
($\hbar/kT  = 2.8$ ps). To explain this surprising longevity, 
we  develop  an  analytical model for the time evolution of the
coherences,   based   on  Eq.~(\ref{coh}).  The  model  provides
physical  insight  into  the  role  of  atomic  energy  levels,
transition  dipole  moments,  and  the  coherence  time  of the
radiation,  as  they  determine  the  coherent evolution of the
Rydberg atom.
In  particular,  the results show coherences that oscillate
  with the frequency determined by the energy level splitting, and
   coherence  properties  of the radiation that enter through the
  ``phase  shifts''  and various prefactors that can be assumed
  to  be  constant in the long time limit ($t\gg \tau_c$).

We emphasize that the results obtained here
apply to the temporal  dynamics  of  any  atomic and/or molecular
  system  coupled  to  incoherent  radiation that is described by an
  arbitrary  stationary correlation function, including CBBR.

%---the only
% requirement  being  that the radiation be stationary, that is,
% describable by a correlation function that depends only on the
% difference of time arguments.
%dynamics the time evolution of the long-lived coherences, it also

Introducing the complex coefficients
 \begin{equation}\label{S}
\MS_{ij}(t) = \mathcal{K}_0^{(+)} (\omega_{i0},t) +\mathcal{K}_0^{(-)}(\omega_{j0},t)
 \end{equation}
and  using  the  property $\MS_{ij}^* = \MS_{ji}$ which follows
from  the definition (\ref{K}), we can rewrite the off-diagonal
density matrix elements [ Eq. (\ref{coh})] as
 \begin{equation}\label{coh2}
\rho_{ij}(t) = \frac{\langle \mu_{i0}\mu_{j0}^* \rangle_p}{\hbar^2}  \frac{1}{\ii\omega_{ij}}  [  \MS_{ij}^* - e^{-\ii \omega_{ij}t} \MS_{ij} ]
 \end{equation}

 The coefficients $\MS_{ij}$  are plotted in Fig. 4 as a function of time for a sample
 pair of eigenstates $|1\rangle = |66s\rangle$ and $|2\rangle = |67s\rangle$
 populated by interaction with CMB starting from the $|65p\rangle$ initial Rydberg
 state (see Fig. 1). The states are separated by an energy gap of
 $\hbar\omega_{21} = 0.86$ cm$^{-1}$ ($1/\omega_{21}=6.2$ ps).
  The  correlation function $\mathcal{C}(\tau)$ of the blackbody
  radiation  decays  on  the timescale $t_c \sim 3\tau_c \approx
  10$  ps  (see Appendix A).
  Accordingly,  both  the  magnitudes  and  the  phases  of  the
  coefficients  $\MS_{ij}  = |\MS_{ij}|e^{\ii \phi_{ij}}$ display
  time-dependent  behavior  during  times $t<t_c$ (here $\sim$10-15 ps),  after  which ($ t > t_c$)  they  can  be  well  approximated  by
  a constant (the  constant $\MS_{ij}$ approximation, see
  Fig.  4). Note that the diagonal matrix elements $\MS_{ii}$
  are real.

For  the  absolute  value  of  the  off-diagonal density matrix
elements in Eq. (\ref{coh2}), we find
 \begin{equation}\label{coh_abs}
|\rho_{ij}(t)| = \frac{| \langle \mu_{i0}\mu^*_{j0} \rangle_p|}{\hbar^2}  \frac{|\MS_{ij}|}{|\omega_{ij}|} 2| \sin (\phi_{ij} - \omega_{ij}t/2)|
 \end{equation}
 whereas the real and imaginary parts of the coherences are given by
 \begin{align}\label{cohReIm}\notag
\text{Re}\rho_{ij}(t) &= -\frac{ \langle \mu_{i0}\mu^*_{j0} \rangle_p}{\hbar^2}  \frac{|\MS_{ij}|}{\omega_{ij}}
\left[ \sin (\phi_{ij} - \omega_{ij}t) + \sin\phi_{ij} \right]; \\
\text{Im}\rho_{ij}(t) &= \frac{ \langle \mu_{i0}\mu^*_{j0} \rangle_p}{\hbar^2}  \frac{|\MS_{ij}|}{\omega_{ij}}
\left[ \cos (\phi_{ij} - \omega_{ij}t) - \cos\phi_{ij} \right];
 \end{align}
  These  expressions  show  that  the  absolute magnitude of the
  coherence   oscillates   with  the  frequency  $\omega_{ij}/2$
  determined   by   the   energy   splitting   between  the  two
  eigenstates.  The  real  and imaginary parts of the coherences
  oscillate  at  twice  this  frequency.  A  related  result was
  obtained in Ref. \cite{Zaheen} for the case of white noise. Equation (\ref{coh_abs}) is, however, 
  more general, as it
  applies to any kind of colored noise described by an arbitrary
  correlation function $C(\tau)$ (the only essential requirement
  being  that  the  noise  is  stationary  so  that Eq. (\ref{coh})
  applies).  Each particular correlation function determines
  the  dynamics  through  different  $\MS_{ij}$  coefficients  in  Eq.
  (\ref{coh2}), which contains the characteristics of the radiation.

 Equations  (\ref{cohReIm})  provide convenient analytic expressions
 for  noise-induced  coherences  in  the limit $t \gg \tau_c $,
 and are straightforward to parametrize via the coefficients
 $\MS_{ij}$.  We note that these expressions could significantly reduce 
 computational challenges in, e.g., calculating the density matrix dynamics
 of molecular systems. 
  Figure  5(a) shows the real part of the coherence
 $\rho_{12}(t)$    calculated    using    Eq.   (\ref{cohReIm})
 parametrized by the constant, asymptotic values for $|\MS_{12}|$
 and  $\phi_{12}$  from  Fig.  4.  The  analytic  result  is in
 excellent   agreement  with  the  exact  calculation,  thereby
 validating   the   constant   $\MS_{ij}$   approximation.  The
 disagreement  at short times is expected, since the $\MS_{ij}$
 vary strongly in this region, and hence cannot be approximated
 by constants. In particular, Eqs.~(\ref{cohReIm}) parametrized
 by  the  asymptotic  values  of  $\MS_{ij}$ disagrees with the
 correct  zero-time  result $\rho_{ij} (0)= 0$. This drawback can be
 remedied, if desired,  by  using  a  different  parametrization  such  that
 $\MS_{ij}(t) \to 0$ as $t\to 0$.

  A  useful measure of the relative importance of coherences and
  populations       is       the       ratio       \cite{Zaheen}
  \begin{equation}\label{Cmeasure}
\mathcal{C}_{ij}(t) = \frac{|\rho_{ij}(t)|}{\rho_{ii}(t) + \rho_{jj}(t)}
 \end{equation}
A  small  value of $\mathcal{C}_{ij}$ indicates that the magnitude of the
coherence   $\rho_{ij}$  is  small  compared  to  that  of  the
populations,   which   is   characteristic  of  a  nearly  pure
statistical  mixture.  Hence, the  timescale  for the decay of $\mathcal{C}$ can
  be  used to quantify the evolution from a purely coherent
state   at   $t=0$  to  a  statistical  mixture  of  stationary
eigenstates.  

 To   obtain   an  analytic  expression  for  the
$\mathcal{C}$-ratio,  we  use an approximate result for state populations
obtained     from     Eq.~(\ref{pop})     by    omitting    the
$\mathcal{K}_1^{(\pm)}$ terms, which are negligible compared to
the  other  two  terms  in  the limit $t\gg \tau_c$ \cite{TBP}.
Combining the resulting expression with Eq. (\ref{coh_abs}), we
find
 \begin{multline}\label{CmeasureA}
%\mathcal{C}_{ij}(t) = \frac{1}{|\omega_{ij}|t}  \frac{|\langle \mu_{i0}\mu_{j0} \rangle_p |}{ \langle |\mu_{i0}|^2 \rangle_p \MS_{ii} + \langle |\mu_{j0}|^2 \rangle_p \MS_{jj}  } \\ \times 2 {|\MS_{ij}|} | \sin (\phi_{ij} - \omega_{ij}t/2)|
\mathcal{C}_{ij}(t) = \frac{1}{|\omega_{ij}|t}  \frac{|\langle \mu_{i0}\mu_{j0} \rangle_p |}{ \langle |\mu_{i0}|^2 \rangle_p \MS_{ii} + \langle |\mu_{j0}|^2 \rangle_p \MS_{jj}  } |\MS_{ij}|   \\ \times 2 |\sin (\phi_{ij} - \omega_{ij}t/2) |
\end{multline}
Figure  5(b)  plots the time variation of the $\mathcal{C}$-ratio for the
Rydberg  states  $|1\rangle$  and $|2\rangle$ defined above. It is seen to  decay in time as $1/(|\omega_{ij}|t)$ and
oscillates  with  the  frequency  $\omega_{ij}/2$,  due  to the
oscillating behavior of the absolute magnitude of the coherence
(\ref{coh_abs}).  This shows that the coherences between the
Rydberg  levels, evident  in Figs. 2 and 3, survive for  long
times  because of the small energy splittings between the levels
populated  by  one-photon absorption and stimulated emission of
CBBR. Longevity of coherences in association with small energy level
splittings has been noted before, albeit in different contexts and
with different functional dependences on the splittings\cite{Elran1,Elran2,Pachonjpcl}
Indeed, in this case, the dependence on $\omega_{ij}t$ is reminiscent of
the energy-time uncertainty principle, as the system strives, in time, 
to perceive individual energy levels.

\section{Summary and future prospects}

In summary,  the long-lived temporal coherence, and associated decoherence, in
Rydberg atoms induced by the sudden turn-on of CBBR at 2.7 K  has been examined.
%, which can be observed experimentally by measuring time-dependent fluorescence from the Rydberg states populated by the interaction with the radiation.
The physical mechanism behind the coherences and their slow decay is the
long coherence time of CBBR and the 
small energy level splittings of the Rydberg levels excited by the CMB.
The large transition dipole moments of the Rydberg atoms
make these coherences manifest in various physical observables.
Directly measuring CMB
coherence properties via fluorescence detection would require 90 dB amplification of the
incident CMB signal, beyond current practice of 67 dB.
At present, achieving such a high gain experimentally over a  broad frequency interval (10-20 GHz) is a formidable challenge.
However, recent
developments in amplification technology  allow for higher gains over much
wider frequency intervals than possible with HEMT amplifiers \cite{NatPhys},
and these may resolve experimental challenges associated with carrying
out the proposed experiment.

\begin{figure}[t]
    \centering
    \includegraphics[width=0.4\textwidth, trim = 0 0 0 0]{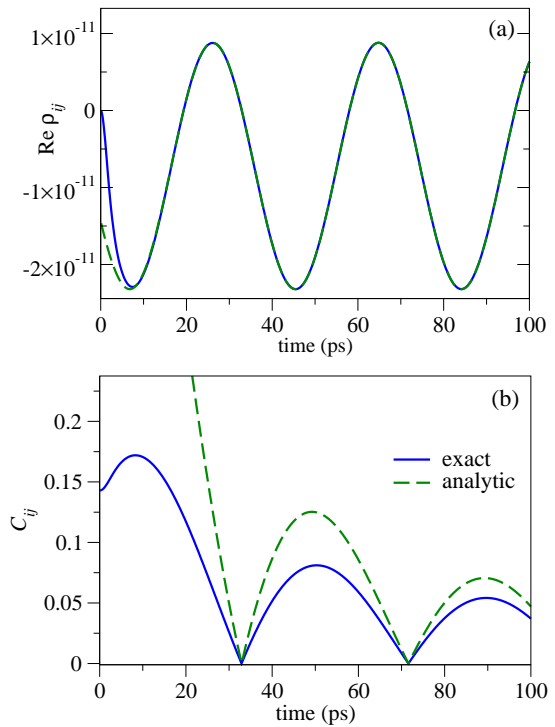}
    \renewcommand{\figurename}{Fig.}
    \caption{(a) The real part of the coherence between levels $|1\rangle = |66s\rangle$ and $|2\rangle = |67s\rangle$ as a function of time. Full line -- exact result (Eq. (5), Fig. 2), dashed line -- constant $\MS_{ij}$ approximation (Eq. \ref{cohReIm}). (b) The ratio $\mathcal{C}_{12} = |\rho_{12}(t)|/(\rho_{11}(t) + \rho_{22}(t))$ as a function of time. Full line -- exact result, dashed line -- constant $\MS_{12}$ approximation, Eq. (\ref{CmeasureA}). }\label{fig:Sij}
\end{figure}

%Such measurements  possibly providing important insights into early Universe cosmology	  \cite{Hakim,Weinberg,Hinshaw,Amanullah}. Initial experimental verification of this approach could be carried out in a cold chamber with walls at 2.7 K, with subsequent experiments aimed at CMB itself.

%We note that Hulet {\it et al.} observed CBBR-induced radiative
%$transfer  from  the  $n=22$  circular  Rydberg state of Na at a
%temperature  as  low  as  6.5  K  \cite{6K}. These experiments,
%however,  were  carried out on a far longer timescale (hundreds
%of  $\mu$s)  than  considered in this work, and were limited to
%the observation of CBBR-induced population transfer rates.

% observed population except that  similar experiment have been conduced
%\begin{figure}[t]
%   \centering
%   \includegraphics[width=0.43\textwidth, trim = 0 0 0 0]{Fig2}
%   \renewcommand{\figurename}{Fig.}
%   \caption{Frequency shift enhancement factors for K-He vs temperature: experimental data from Ref. \cite{Walker05}, present {\it ab initio} calculations (full line). }\label{fig:enhancement_factors}
%\end{figure}

%\begin{figure}[t]
%   \centering
%   \includegraphics[width=0.43\textwidth, trim = 0 0 0 0]{Fig3}
%   \renewcommand{\figurename}{Fig.}
%   \caption{Rate constants for spin exchange in K-$^3$He collisions vs temperature at $B=1$ G: $k_\alpha$ (dashed line), $k=k_\alpha+k_\beta$ (full line); symbol -- experimental result \cite{Walker10}. Also shown (red line) is the calculated $k(T)$ for Ag-He.}\label{fig:rates}
%\end{figure}

Finally, we note  that the  long-lived coherences shown in Fig. 2
can also be observed  with any experimental technique that is
sensitive to coherent superpositions of the atom's excited
states. Examples include  selective field ionization \cite{RA},
photoionization \cite{RA}, and half-cycle  pulse ionization \cite{Jones}.
The former technique also provides a direct route to measuring
non-Markovian deviations from the linear behavior of state populations at
short times (Fig. 2a), which also relates  to the coherence properties
of CBBR \cite{TBP}.

One extension of this work is readily motivated. The study in this paper
has examined the sudden turn-on
associated with a single state prepared in an excited Rydberg state. However, 
slower preparation of Rydberg states,
e.g., using a 15 ps laser pulse is expected to produce \cite{RydbergWP} additional
interesting results. That is, such a pulse prepares a coherent superposition of
five  eigenstates  centered  around  $n=65$,  rather than a single
state,  as  assumed above. This superposition will then couple,
via the CMB, to adjacent $s$ and $p$ Rydberg states. Fluorescence from this
collection of levels is then expected to display
a more complicated pattern of quantum beats than described above,
which then decoheres in time. In addition, since the initial state is then
a prepared superposition of energy eigenstates, decay of decoherence on
assorted time scales is also anticipated \cite{Elran1,Elran2}. Further,
one can consider modifying the laser pulse shape in order to enhance the
quantum beat signal. Such studies are underway.

%Such an analysis is now underway and will be presented in future work \cite{TBP}.

%The latter technique, which directly measures the momentum distribution of
%the Rydberg electron \cite{Jones} may be particularly well-suited
%for the purpose of observing CBBR-induced coherences due to the
%sensitivity of the momentum distribution to quantum interference
%effects \cite{Robicheaux,TBP}.

% ab include  Rydberg atom detection that is suitable  that is sensitive in other detection schemes: SFI, photoionization, etc.. Many-body effects for enhancing observed fluorescence intensity  ($N^2$ scaling is probably not unfeasible given large wavelengths of CMB)

%Applications to astrophysics (early Universe, coheence effects have never been explored as a probe into its dynamics, although suggested by Hakim). More generally, what is the role played by these coherence effects in molecular excitation, etc... speculations on astrochemistry...

\section{Acknowledgements}
We thank Dr. Marian Pospieszalski,  Dr. Hossein Sadeghpour, Prof. John Polanyi,  Dr. Colin
Connoly and Dr. Leonardo Pach{\'o}n
for discussions. This work was supported by the Natural
Sciences and Engineering Research Council of Canada and the U.S.
Air Force Office of Scientific Research under contract number
FA9550-13-1-0005.

\section{Appendix}

 This Appendix  outlines the derivation of the equations of
motion  for  the  density  matrix [Eqs. (3) and (5)] that
describe  the  interaction of a Rydberg atom with 
blackbody radiation.

The  time  evolution  of  the density matrix for a Rydberg atom
interacting with CBBR is given by Eq. (1). For
a    stationary   CBBR   source,   the   correlation   function
$\mathcal{C}(\tau',\tau'')$      is      a      function     of
$\tau=\tau'-\tau''$  only.  The absolute value and the phase of
the CBBR correlation function given by Eq. (2)
are plotted in Fig. 6 as a function of $\tau$ for $T_\text{CMB}
= 2.718$ K \cite{KanoWolf}.

By  changing  the  integration  variables  $\tau_\pm=  \tau'\pm
\tau''$, Eq. (1) can be recast in the form
\begin{multline}\label{EOM2}
\rho_{ij}(t)  = \frac{\langle \mu_{i0}\mu_{j0}^* \rangle_p}{\hbar^2}  e^{-\ii\omega_{ij}t}  \frac{1}{2} \biggl{[} \int_{-t}^0 d\tau_- \int_{-\tau_-}^{\tau_- + 2t} d\tau_+ f(\tau_+,\tau_-)
\\ + \int_{0}^{t} d\tau_- \int_{\tau_-}^{2t-\tau_-} d\tau_+ f(\tau_+,\tau_-)\biggr{]},
\end{multline}
where $f(\tau_+,\tau_-)= \mathcal{C}(\tau_-) e^{\ii\omega_{i0}(\tau_++\tau_-)/2} e^{-\ii\omega_{j0}(\tau_+-\tau_-)/2}$. For $i=j$, the integrand simplifies to
\begin{equation}\label{fii}
f(\tau_-)=\mathcal{C}(\tau_-) e^{\ii\omega_{i0}\tau_-}
\end{equation}
allowing the integration
 over  $\tau_+$ in Eq. (\ref{EOM2}) to be performed analytically
 to yield the population dynamics
\begin{equation}\label{populations}
\rho_{ii}(t)  = \frac{\langle |\mu_{i0}|^2 \rangle_p}{\hbar^2}  \left[ t\mathcal{I}_0(t)  + \mathcal{I}_1(t) \right]
\end{equation}
with
\begin{align}\label{I}
\mathcal{I}_0 (t) &= \int_{-t}^t f(\tau_-)d\tau_-; \\
\mathcal{I}_1(t) &= \int_{-t}^0 \tau_- [f(\tau_-) + f(-\tau_-)]d\tau_-
\end{align}
Splitting  the  range  of  integration in the first term on the
right-hand  side  into  positive and negative $\tau_-$ regions,
relabeling  the  integration  variable  $\tau_-  \to \tau$, and
using Eq.~(\ref{fii}) we find
\begin{equation}\label{I0}
\mathcal{I}_0 (t) = \int_{0}^t [\mathcal{C}(\tau) e^{\ii\omega_{i0}\tau} + \mathcal{C}(-\tau) e^{-\ii\omega_{i0}\tau}] d\tau
\end{equation}
and
\begin{equation}\label{I1}
\mathcal{I}_1 (t) = -\int_{0}^t \tau [\mathcal{C}(\tau) e^{\ii\omega_{i0}\tau} + \mathcal{C}(-\tau) e^{-\ii\omega_{i0}\tau}] d\tau.
\end{equation}
Introducing the half-Fourier transforms
\begin{align}\label{hft}
\mathcal{K}^{(\pm)}_0 (\omega,t) &= \int_{0}^t \mathcal{C(\pm\tau)} e^{\pm \ii\omega\tau} d\tau \\
\mathcal{K}^{(\pm)}_1(\omega,t) &= \int_{0}^t \tau \mathcal{C(\pm\tau)} e^{\pm \ii\omega\tau} d\tau,
\end{align}
we obtain Eq. (3) in the text.

\begin{figure}[t]
	\centering
	\includegraphics[width=0.37\textwidth, trim = 0 0 0 0]{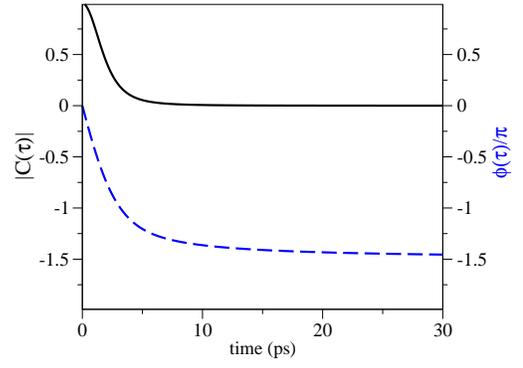}
	\renewcommand{\figurename}{Fig.}
	\caption{Time dependence of CBBR correlation function $\mathcal{C}(\tau) = |\mathcal{C}(\tau)|e^{\ii\phi(\tau)}$ for $T = 2.718$ K (in units of $\mathcal{E}_0^2$, see Eq.~ (2) of the main text). Full line -- absolute magnitude ($|\mathcal{C}(\tau)|$), dashed line -- phase ($\phi(\tau)/\pi$) \cite{KanoWolf}.}\label{fig:scaling_rates}
\end{figure}

In  the  case of $i\ne j$ (off-diagonal elements of the density
matrix),  the  integrand  depends on both $\tau_+$ and $\tau_-$
via
\begin{equation}\label{coher}
f(\tau_+,\tau_-)= \mathcal{C}(\tau_-) e^{\ii\tau_+(\omega_{i0}-\omega_{j0})/2} e^{\ii\tau_-(\omega_{i0}+\omega_{j0})/2}
\end{equation}
Substituting Eq. (\ref{coher}) in Eq. (\ref{EOM2}) and evaluating the integral over $\tau_+$ analytically (which is straightforward since $\mathcal{C}(\tau_-)$ is a function of $\tau_-$ only), we arrive at the result
\begin{multline}\label{coher2}
\rho_{ij}(t)  = \frac{\langle \mu_{i0}\mu_{j0}^* \rangle_p}{\hbar^2 (\ii\omega_{ij})} \biggl{\{} \int_0^t \left[ \mathcal{C}(\tau)e^{\ii\omega_{j0}\tau} +\mathcal{C}(-\tau)e^{-\ii\omega_{i0}\tau}\right] d\tau  \\
- e^{-\ii\omega_{ij}t} \int_0^t \left[ \mathcal{C}(\tau)e^{\ii\omega_{i0}\tau} +\mathcal{C}(-\tau)e^{-\ii\omega_{j0}\tau}   \right]d\tau   \biggr{\}}
\end{multline}
With the help of the definition (\ref{hft}), we obtain Eq. (5) in the above text.

 \begin{equation}\label{zetaf}
\zeta(4,a) = \frac{1}{\Gamma(4)} \int_{0}^\infty \frac{x^3 e^{-ax}}{1-e^{-x}} dx,
\end{equation}
where $\Gamma(x)$ is a Gamma function, we get
\begin{multline}\label{limit}
\rho_{ii}(t\to \infty)  = \frac{\langle |\mu_{i0}|^2  \rangle_p}{\hbar^2}  \int_{-\infty}^\infty \mathcal{C}(\tau) e^{\ii\omega_{i0}\tau}d\tau  \\
= \mathcal{E}_0^2 \frac{90}{\pi^4 \Gamma(4)}\int_{0}^\infty d\omega \frac{\omega^3 e^{-(1+\ii\lambda\tau)\omega}}{1-e^{-\omega}} \int_{-\infty}^\infty d\tau e^{-\ii\lambda\tau \omega}e^{\ii\omega_{i0}\tau}
\end{multline}
The integral over $\tau$ is readily evaluated in terms of the Dirac $\delta$-function
%\begin{equation}\label{delta}
($\int_{-\infty}^\infty e^{-\ii(\lambda \omega-\omega_{i0})\tau} d\tau = 2\pi \delta(\lambda\omega - \omega_{i0}).$)
%\end{equation}
and Eq. (\ref{limit}) reduces to the Fermi Golden Rule result given by Eq. (4)
in the text ($\lambda = kT/\hbar$)
\begin{align}\label{limit2}\notag
\rho_{ii}(t\to \infty)  &= \mathcal{E}_0^2 \langle |\mu_{i0}|^2\rangle_p \frac{90}{\pi^4} \frac{2\pi}{\Gamma(4)} \frac{1}{\lambda^4} \frac{\omega_{i0}^3}{ e^{\omega_{i0}/\lambda} -1 } t \\
& = \frac{4\langle |\mu_{i0}|^2\rangle_p}{3\hbar c^3}\frac{\omega_{i0}^3}{ e^{\hbar\omega_{i0}/kT} - 1 } t
\end{align}
Note that the proportionality coefficient in the second line of
Eq.  (\ref{limit2}) is the BBR-induced transition rate $W_{0\to
i}$.

%\newpage

\end{document}